# PECULARITIES OF COSMIC RAY MODULATION IN THE SOLAR MINIMUM 23/24.

M. V. Alania[1,2], R. Modzelewska[1] and A. Wawrzynczak[3]

[1] Institute Math. & Physics, Siedlce University, 3 Maja 54, 08-110 Siedlce, Poland.
[2] Institute of Geophysics, Tbilisi State University, Tbilisi, Georgia.
[3] Institute of Computer Science, Siedlce University, 3 Maja 54, 08-110 Siedlce, Poland.
alania@uph.edu.pl, renatam@uph.edu.pl, awawrzynczak@uph.edu.pl,

**Abstract**

We study changes of the galactic cosmic ray (GCR) intensity for the ending period of the solar cycle 23 and the beginning of the solar cycle 24 using neutron monitors experimental data. We show that an increase of the GCR intensity in 2009 is generally related with decrease of the solar wind velocity U, the strength B of the interplanetary magnetic field (IMF), and the drift in negative (A<0) polarity epoch. We present that temporal changes of rigidity dependence of the GCR intensity variation before reaching maximum level in 2009 and after it, do not noticeably differ from each other. The rigidity spectrum of the GCR intensity variations calculated based on neutron monitors data (for rigidities > 10 GV) is hard in the minimum and near minimum epoch. We do not recognize any non-ordinary changes in the physical mechanism of modulation of the GCR intensity in the rigidity range of GCR particles to which neutron monitors respond.
We compose 2-D non stationary model of transport equation to describe variations of the GCR intensity for 1996-2012 including the A>0 (1996-2001) and the A<0 (2002-2012) periods; diffusion coefficient of cosmic rays for rigidity 10-15 GV is increased by ~ 30% in 2009 (A<0) comparing with 1996 (A>0). We believe that the proposed model is relatively realistic and obtained results are satisfactorily compatible with neutron monitors data.

## 1. Introduction

An elongated solar cycle 23 caused big interest to study features of the ending part of the solar cycle 23 and the beginning of the solar cycle 24 (further in this paper called 'solar minimum 23/24') based on direct observations of processes taking place in interior of the Sun, solar atmosphere and heliosphere. In this introduction we shortly analyze results of recent publications dealing with solar cycle 23, and for comparison solar cycle 22, as well.

Dikpati [2013] demonstrated that during solar cycles 22 and 23 the meridional circulation pattern on the Sun behaved distinctly differently. Particularly, the Sun's surface plasma flow was poleward all the way up to the pole persisted during the major part of solar cycle 23, while during solar cycle 22 the poleward surface flow ended at latitude ~ 60°. The distinction between solar cycles 22 and 23 was ascribed to the difference in the structures of the coronal holes for cycles 22 (1986-1996) and 23 (1996-2008); particularly a large, stable coronal hole structure was observed during cycle 22, but not- during cycle 23 [Dunzlaff et al.,2008].
The most probable source of the differences between solar cycles 22 and 23 is the dynamic of the Sun's interior which, through the turbulent dynamo process, influences on the generation and evolution of the Sun's global magnetic fields [Parker, 1994]. The evolution of the interplanetary magnetic field (IMF) in the heliosphere in the anomalously long solar minimum 23/24 shows the weakest magnetic field strengths so far observed during the space age [Schwadron *et al.*, 2010]. The solar minimum 23/24 (A<0) is different from the previous minimum 22/23 (A>0). Because of very peculiar character of the solar minimum 23/24, modulation of the galactic cosmic ray (GCR) intensity in this period was broadly investigated [*e.g.*, Heber *et al.*, 2009; McDonald *et al.*, 2010; Schwadron *et al.*, 2010; Moraal and Stoker, 2010; Mewaldt *et al.,* 2010]. Nevertheless, the fundamental physical processes that accelerate the solar wind have not changed. So, one can state that the minimum 23/24 is not particularly unusual [Zhao and Fisk, 2011]. GCR modulation in relation to solar activity indexes and heliospheric parameters in the minimum 23/24 was recently studied by Paouris *et al.* [2012]. Chowdhury et al. [2013] showed that GCR intensity experience various types of modulation from different solar activity features and influence on space weather and the terrestrial climate.

In this paper we focus our attention to: (1) step like changes of the GCR intensity in unusual elongated descending epoch (A<0) of solar cycle 23 (2004-2009), (2) roles of decrease of solar wind velocity U and strength B of the IMF, and drift effect in the increase of the GCR intensity in 2009, (3) temporal changes of rigidity spectrum of the GCR intensity variations before and after reaching the maximum of GCR intensity in 2009, and (4) 2-D non stationary modeling of GCR transport in heliosphere for period of 1996-2012 including two opposite polarity periods 1996-2003 (A>0) and 2004-2012 (A<0).

**Experimental data**

Fig. 1 presents changes of the 27-day averages of the GCR intensity based on Oulu NM data calculated vs. average GCR intensity in March-July 2009 and recalculated to the free space using coupling coefficients [Yasue *et al.*, 1982], solar wind velocity U, IMF's magnitude B and tilt angle (TA) of the heliospheric neutral sheet (HNS) for 1996-2012. To clearly visualize the rigidity dependence of the GCR intensity variations, in 2003-2012, Fig. 2 presents the temporal changes of the smoothed over 3 month data for neutron monitors (NMs) with different cut off rigidities ($R_c$): Oulu ($R_c$=0.81 GV), Hermanus ($R_c$=4.9 GV), and Potchefstroom ($R_c$=7.3 GV). There are plainly visible increases of the GCR intensity in 11/2003-5/2004, 8/2005-3/2006, 1/2007-5/2007 and 6/2008-5/2009, each of them finishing with the plateaus lasting ~6-8 months. These features can be associated with the creation of the almost constant homogeneous electromagnetic conditions in the restricted local vicinity of the heliosphere, where a formation of the impulsive step-like changes of the GCR intensity takes place. Consequently, the heliosphere in the considered period can be regarded as the structure of concentric sphere-like zones with homogeneous, relatively constant electro-magnetic conditions (intensity of GCR in scope of each sphere-like zone remains almost constant). These unique step-like changes of the GCR intensity in 2003-2009 during descending 11- year solar cycle needs a detail study in relation with a mechanism of 11-year variation of cosmic rays [e.g. Lockwood and Webber, 1996]. One can note that step-like changes of the GCR intensity (lasting ~ 6-8 months) are observed preferentially in the A<0 polarity epochs, when the MIRs and GMIRs type structure are preferentially observed in the heliosphere [e.g., Burlaga and Ness, 2000].

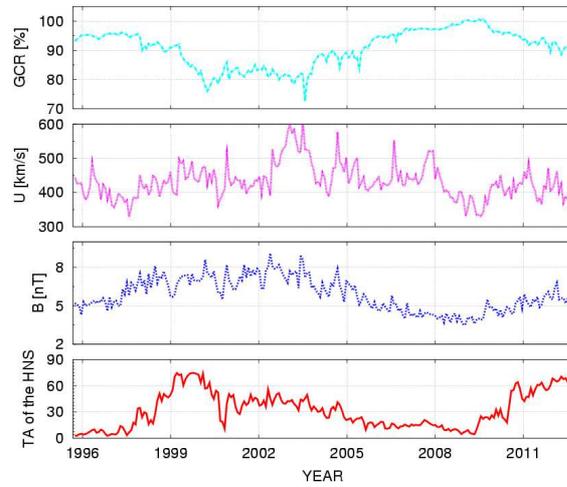

Fig. 1. Temporal changes of the 27-day averages of the GCR intensity in free space obtained based on Oulu NM data, solar wind velocity U, magnitude B of the IMF and TA for 1996-2012.

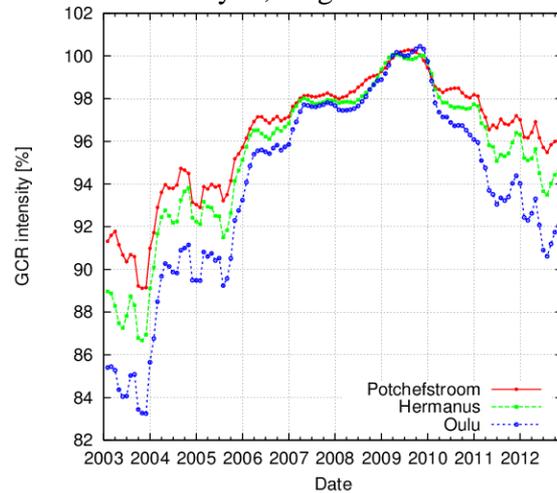

Fig. 2. Temporal changes of the GCR intensity smoothed over 3 month for Oulu, Hermanus and Potchefstroom NMs for 2003-2012 calculated vs. average GCR intensity in March-July 2009.

To study the nature of the GCR intensity increase in 2009 in detail, we consider relatively short period 2008-2010. Fig. 3 (top panel) presents time lines of the GCR intensity I by Oulu NM and solar wind velocity U, and changes of the correlation coefficient between them for different delay times in unites of Bartel's Rotation (BR). Fig. 3 (bottom panel) shows an existence of the anti-correlation between I & U (r=-0.68±0.09) with the maximal magnitude of r for zero delay time in BR. This result we partially account for the reducing of GCR convection stream owing to the failing of the solar wind velocity. Fig. 4 presents: (top panel) temporal changes of the 27-day averages of the GCR intensity for Oulu NM (red points) and IMF's magnitude B (green squares), and shifted changes of B (blue stars) for 3 BRs with respect to GCR intensity; and (bottom panel) changes of the correlation coefficient between GCR intensity by Oulu NM and B for different delay times in 2008-2010. It is clearly seen that there exists an anti-correlation between I & B (r=-0.73±0.08) for the delay time ~ 3BR. Due to inverse dependence of diffusion coefficient of cosmic rays on B, a dropping of B cause an increase of diffusion coefficient and, consequently, an appearance of additional diffusion stream of cosmic rays. Fig. 5 presents: (top panel) temporal changes of the 27-day averages of the GCR intensity for Oulu NM (red points) and TA (green squares), and TA shifted for 5 BRs (blue stars) in 2008-2010. One can note a strong inverse correlation between I & TA with correlation coefficient, r=-0.92±0.05 for the delay time 5 BRs (Fig. 5, bottom panel). Considering 5 BRs as an average time-interval determining a distance ~35-40 AU covered by solar wind (with U ~ 400 km/s), we can assume that processes in this vicinity of the space are generally responsible for increase of the GCR intensity in 2009. On the other, a decrease of the TA during 22-24 months (2008-2009) due to gradually enhanced drift effect has created the additional stream of GCR contributing to increase of the GCR intensity in 2009. Then, at the end of 2009 an enhancement of TA caused corresponding reduces of the GCR intensity and played an important role in formation of the distinctive impulsive shape of the GCR intensity in 2009. Beside the direct roles of diminishing U and B in the increasing of I, the maximum intensity I in 2009 is established by drift effect with delay time 5 BRs between I and TA in negative polarity (A<0) period; for A<0 polarity epoch a maximum I of the GCR intensity in heliosphere is observed immediate proximity to the HNS's location. So, higher intensity of GCR at the earth orbit corresponds to the location of HNS in ecliptic region, while lower level of the GCR intensity- to location of HNS out of ecliptic region. Thus, the increase of the GCR intensity in 2009 is carried out on the background of 0 BR delay time between GCR intensity I and diminishing solar wind velocity U, ~ 3BR delay time between I and diminishing IMF strength B, and ~5BR delay time between I and TA.

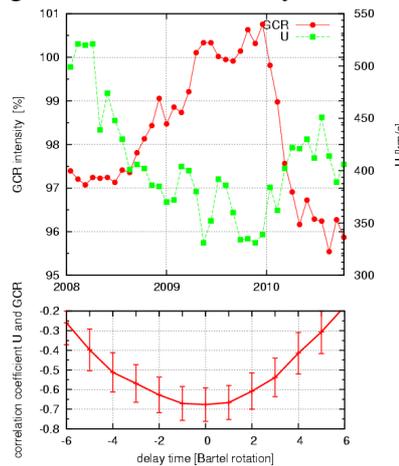

Fig. 3. (top panel) Temporal changes of the 27-day averages of the GCR intensity for Oulu NM and solar wind velocity U; (bottom panel) correlation coefficient between GCR and U for different delay times in 2008-2010.

We assume that the delay time ~5 BR between I and TA determines the size of local region (~ 35-40 AU) where, the process of increase of the GCR intensity in 2009 took place. So, we state that the increase of the GCR intensity in 2009 is generally related with diminishing of solar wind velocity U (decrease of convection), diminishing of the IMF's strength B (increase of diffusion coefficient), and by clearly seen drift effect associated with the changes of TA in A<0 period.

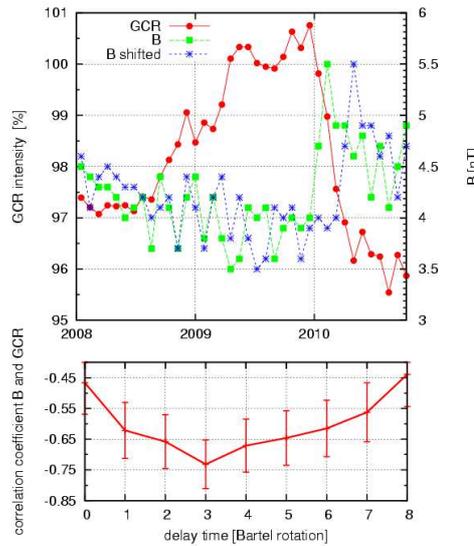

Fig. 4. (top panel) Temporal changes of the 27-day averages of the GCR intensity for Oulu NM (red points) and IMF strength B for 0 delay time (green squares) and shifted for 3 BRs (blue stars); (bottom panel) correlation coefficient between GCR and IMF for different delay times for 2008-2010.

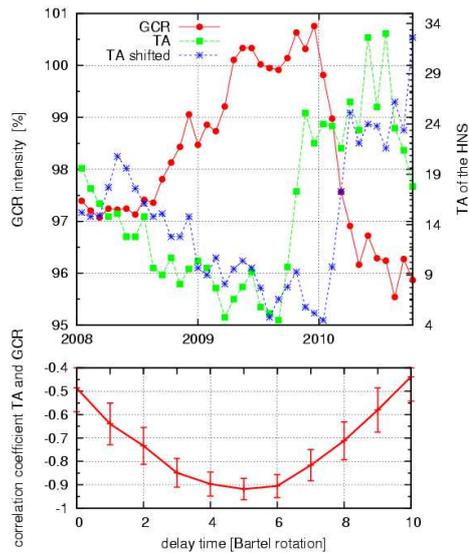

Fig. 5. (top panel) Temporal changes of the 27-day averages of the GCR intensity for Oulu NM (red points) and TA for 0 delay time (green squares) and TA shifted for 5 BRs (blue stars); (bottom panel) correlation coefficient between changes of the GCR intensity and TA for different delay times in 2008-2010.

## 2. Rigidity spectrum of the GCR intensity variations around the solar minimum in 2009

The changes in the primary rigidity spectrum of the GCR modulation around the last minima of solar activity were studied by various authors. Adriani et. al. [2013] presented the galactic proton spectra measured by PAMELA from July 2006 up to December 2009. They showed that the primary spectrum of protons with energy ~100 MeV becomes soft as the minimum of solar activity is reached. Bazilevskaya et al. [2012] estimated the spectrum of CR based on GCR intensity observations on balloons and on the ground-based neutron monitors data. This analysis showed that between mid of 2008 and mid of 2009 the primary CR spectrum at the Earth's orbit became softer due to prevailing contribution of particles with energies below several GeV. We would like to underline that in this paper we do not estimate the modulated primary GCR spectrum around the 2009 solar minima, but we are focusing on the energy spectrum of the GCR variation in period of 2006-2012 (Fig. 6). Fig.6 (in arbitrary scales) shows changes of primary spectra in free space, D(R) is modulated and $D_0(R)$ unmodulated primary spectra.

Variations of the primary spectrum of GCR in the interplanetary space are defined as: $\left(\frac{\delta D(R)}{D(R)}\right)_k = \frac{D_k(R) - D_0(R)}{D_0(R)}$, where $D_k(R)$ is the primary spectrum for the 'k' month and $D_0(R)$ - primary spectrum during the reference period (here March-July 2009). Unfortunately, we do not know neither $D_0(R)$ nor $D_k(R)$. However, we can calculate the rigidity spectrum of the GCR variations (see Fig. 6).

A relation between the count rate $N^k$ of a ground-based detector and primary spectrum of GCR at the Earth orbit $D_k(R)$ via yeld function is as follows: $N^k = \int D_k(R) \cdot Y(R) \cdot dR$. So, for any kind of isotropic intensity variations of GCR we can write (eg. Dorman 1963) $J_i^k = (N_i^k - N_i^0)/N_i^0 = \int_{R_i}^{R_{max}} (\delta D(R)/D(R))_k W_i(R, h_i) dR$, where $J_i^k$ is amplitude measured by 'i' detector with the geomagnetic cut off rigidity $R_i$ and the average atmospheric depth $h_i$; $(\delta D(R)/D(R))_k$ is the rigidity spectrum of the GCR intensity variation for the 'k' month, $W_i(R, h_i) = (D_0(R) * Y(R, h_i))/N_i^0$ is the coupling coefficient for 'i' detector.

We assume that (as it is generally adopted and is considered as a good enough approximation) the rigidity R spectrum $\delta D(R)/D(R)$ of the GCR variations, in particular of the 11-year variation, is power law type and can be written as follows [e.g. Dorman 2004, Wawrzynczak and Alania 2010a, Alania and Wawrzynczak, 2012]:

$$\frac{\delta D(R)}{D(R)} = \begin{cases} A\left(\frac{R}{R_0}\right)^{-\gamma} & \text{for} \quad R \leq R_{max} \\ 0 & \text{for} \quad R > R_{max} \end{cases}, \quad (1)$$

where $R_{max}$ is rigidity beyond which the modulation of GCR intensity is not observed and A is a constant quantitatively equals $\frac{\delta D(R)}{D(R)}$ for rigidity R = $R_0$.

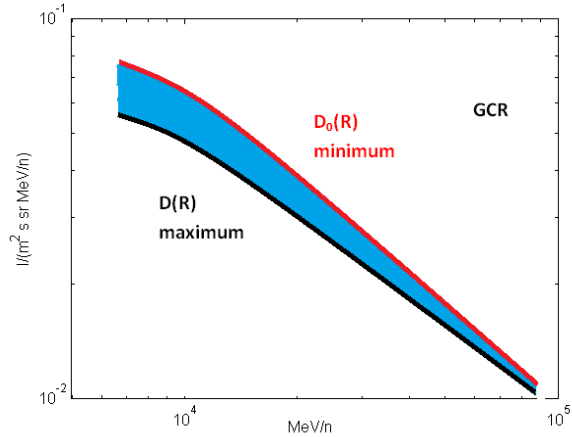

Fig. 6. The change of the arbitrary primary spectra in free space, D(R) is modulated and $D_0(R)$ unmodulated primary spectra. In this paper is calculated the rigidity spectrum of the GCR variation in the energy range ~10-50 GeV. i.e. the spectrum of the difference between $D_0(R)$ - unmodulated and D(R)-modulated primary spectra marked at figure by blue color.

To calculate rigidity spectrum of the GCR intensity variations around the minimum of solar activity in 2009, the amplitudes of the GCR intensity variations for NMs with different cut off rigidities were found. To obtain a statistically reliable value of amplitudes, we took 13 months moving averages for the period of January 2006- April 2013. For calculations were used data of 10 NMs: Apatity ($R_c$=0.65 GV), Cape Shmidt ($R_c$=0.45 GV), Fort Smith ($R_c$=0 GV), Hermanus ($R_c$=4.9 GV), McMurdo ($R_c$=0.01 GV), Moscow ($R_c$=2.46 GV), Oulu ($R_c$=0.81 GV), Potchefstroom ($R_c$=7.3 GV), Rome ($R_c$=6.32 GV) and Thule ($R_c$=0 GV). Results of smoothing data for Potchefstroom, Hermanus, Thule and McMurdo NMs present Fig. 7. The amplitudes for $i^{th}$ NM were calculated as follows: $J_i^k = (N_i^k - N_i^0)/N_i^0$, where $N_i^k$ is the running monthly average count rate (k =1, 2, 3,….month) and $N_i^0$ is the 5 months average count rate in the solar activity minimum i.e. March-July 2009.

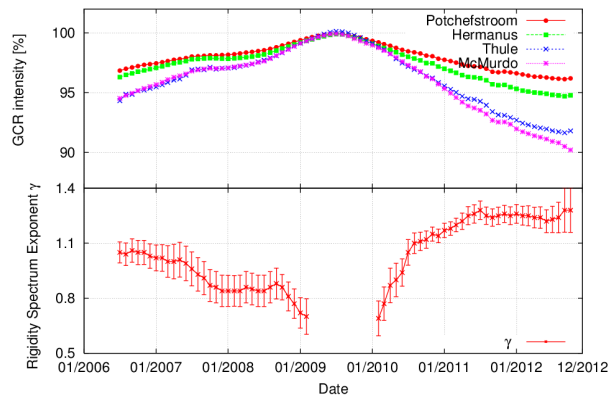

Fig. 7. Timelines of GCR intensity smoothed over 13 months (upper panel) for Potchefstroom, Hermanus, Thule, McMurdo NMs calculated vs. average GCR intensity in March-July 2009 and exponent $\gamma$ (bottom panel) for July 2006 – November 2012.

Based on the method in detail described in [Wawrzynczak and Alania, 2010a; Alania and Wawrzynczak, 2012] we calculated the rigidity spectrum exponent $\gamma$ of GCR variation around the recent solar minimum 23/24 for each month of period 2006-2012 (Fig. 7). Fig. 7 (bottom panel) shows that the exponent γ before reaching point of the maximum intensity in 2009 is gradually diminishing (from γ=1± 0.07 in 2007 to γ=0.77 ± 0.08 at the end of 2008) demonstrating a good tendency of hardening of rigidity spectrum of the GCR intensity variations. After the point of 2009 the exponent γ gradually increases (from γ= 0.69 ± 0.09 at the beginning of 2010, up to γ=1.28± 0.12 at the end of 2012). We can conclude that near minimum epoch of solar activity in 2009, for ascending and descending sides of GCR intensity, a rigidity spectrum of the GCR variation is hard as was obtained for previous solar cycles by Alania et al. [2008, 2010]. Our results of temporal changes of the rigidity spectrum of the GCR intensity variation observed by NMs before and after 2009 show that there does not take place any new physical processes in modulation of cosmic rays during the 23/24 minimum epoch of solar activity comparing with 21/22 and 22/23 solar minima.

## 4. Model of GCR transport in 1996-2012

Our aim in this paper is to account for changes of the GCR intensity in the minimum and near minimum phases of solar cycle 23/24. However, in order to compare results for different phases of solar activity we model the longer period 1996 – 2012. According to specific problem of modeling, we use the Parker's non-stationary transport equation [Parker, 1965]:

$$\frac{\partial N}{\partial t} = \nabla \cdot \left(K_{ij}^S \cdot \nabla N\right) - \left(v_d + U\right) \cdot \nabla N + \frac{1}{3}\frac{\partial}{\partial R}(NR)\nabla U \quad (2)$$

Where $N$ and $R$ are omnidirectional distribution function and rigidity of GCR particles, respectively; $t$ - time, $U$ – solar wind velocity, $v_d$ is the drift velocity.

We model the GCR intensity variations during the #2219-#2447 BR periods i.e. during 218 BRs. A size of the modulation region $r_0 = 100$ AU, the local interstellar medium spectrum presented in Webber and Lockwood, (2001) and Caballero-Lopez and Moraal, (2004) was applied. The anisotropic diffusion tensor of GCR $K_{ij} = K_{ij}^{(S)} + K_{ij}^{(A)}$ consists of the symmetric $K_{ij}^{(S)}$ and $K_{ij}^{(A)}$ – antysymmetric parts. We implement a drift velocity of GCR particles as, $<v_{D,i}> = \frac{\partial K_{ij}^{(A)}}{\partial x_j}$ [Jokipii et al., 1977]. This expression is equivalent to the standard formula for $<v_D>$ [Rossi and Olbert, 1970]. The heliospheric magnetic field vector $\vec{B}$ is taken, as [Jokipii and Kopriva, 1979; Kota and Jokipii, 1983] $\vec{B} = (1 - 2H(\theta - \theta'))\left(B_r \vec{e}_r + B_\varphi \vec{e}_\varphi\right)$ where $H$ is the Heaviside step function changing the sign of the global magnetic field in each hemisphere and $\theta'$ corresponds to the heliolatitudinal position of the HNS, $e_r$ and $e_\varphi$ are the unite vectors directed along the components $B_r$ and $B_\varphi$ of the IMF for the two dimensional Parker field [Parker, 1958]. Parker's spiral heliospheric magnetic field is implemented through the angle $\psi = \arctan(-B_\varphi/B_r) = \arctan(\Omega r \sin\theta/U)$ in anisotropic diffusion tensor of GCR particles ($\psi$ is the angle between magnetic field lines and radial direction in the equatorial plane) and ratios

$\beta = K_\perp/K_{II}$ and $\beta_1 = K_d/K_{II}$ of the perpendicular $K_\perp$ and drift $K_d$ diffusion coefficients to the parallel $K_\parallel$ diffusion coefficient: $\beta = 1/(1+\omega^2\tau^2), \beta_1 = \omega\tau/(1+\omega^2\tau^2)$, where $\omega\tau = 300 B\lambda R^{-1}$, with a mean free path $\lambda$ of GCR. So, for the cosmic ray particles of rigidity R >10GV the perpendicular $K_\perp$ and drift $K_d$ diffusion coefficients are proportional to the parallel $K_\parallel$ diffusion coefficient. The billiard ball type diffusion is not generally the best approximation [Parhi et al. 2004; Shalchi et al., 2010] but it works well at high rigidities R >10 GV to which NMs respond [Jokipii, 1971; Shalchi, 2009].

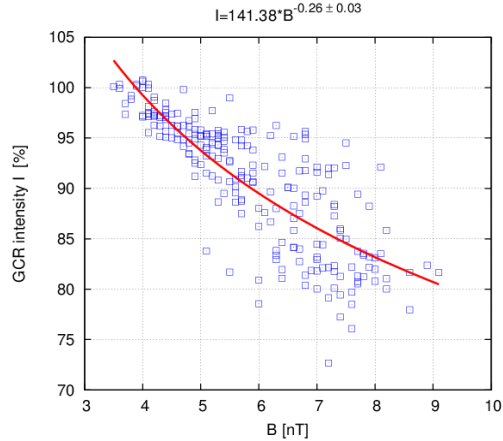

Fig. 8. Scatter plot of the 27-day averages of GCR intensity I and magnitude B of the IMF for 1996-2012.

Unfortunately, to compose a relatively realistic model of GCR propagation describing long period variations in the heliosphere is not easy due to necessity of setting up a few free modulation parameters. However, we used comparatively measurable and estimable parameters, among them: (1) the 27-day averages of the in situ measurements of IMF magnitude $B_0$, (2) solar wind velocity U and (3) TA (Fig. 1). The solar wind velocity U is responsible for the convection, and TA determines a character of drift of cosmic ray particles in heliosphere through drift velocity in A<0 epoch. The B influence on propagation of cosmic rays through diffusion coefficient K, so we assume that, $K\propto B^\alpha$, where α is unknown and should be found from experimental data. To take into account a dependence of the GCR intensity I upon diffusion coefficient K ($I\propto K$) and a dependence of the latter upon $B^\alpha$ ($K\propto B^\alpha$), we have built a scattering diagram of the 27-day averages of I upon B in 1996-2012 (Fig. 8). The scattering diagram presented in Fig. 8 shows that α=-0.26±0.03; we believe that a dependence of I upon B results in some degree an inclusion of delay time between I and B to the Parker's transport equation.

A parallel diffusion coefficient used in modeling is expressed, as:

$$K_{II} = K_0 K(r) K(B_0) K(t) K(R, \gamma(t, R)), \qquad (3)$$

where

- $K_0 = 1.05\times10^{21} cm^2/s$,
- $K(r) = 1 + 0.5 r/r_0$;
- $K(B_0) = B_0^{-0.26}$, obtained based on data analysis in the period of 1996-2012 (Fig.8), $B_0$ is magnitude of the IMF at the Earth's orbit.

- $K(R, \gamma(t,R)) = R^{\gamma(t,R)}$, contributes to the changes of the parallel diffusion coefficient $K_{II}$ due to dependence on the GCR particles rigidity R and is taken as
  $\gamma(t,R) = 0.7 * R^{0.1}$ in period of 1996 and then changes as
  $\gamma(t,R) = (7.9252*t^4 - 13.33*t^3 + 5.8787*t^2 - 0.4797*t + 0.7) * R^{0.1}$ .
- $K(t) = Exp(4.2*(2 - \gamma(t,R)))$, function introduced to make a consistent changes of diffusion coefficient $K_{II}$ throughout the 11–year cycle of solar activity.

The exponent $\gamma(t,R)$ reflects the changes of the rigidity spectrum of the GCRs. According to the experimental data analysis [e.g., Alania et al. 2008, and references therein] this exponent increases in the maximum of solar activity. In this paper we assume that γ(t,R) reflects this changes i.e. in minimum 1996 of solar activity γ ≈ 0.7, in the maximum it increases up to γ ≈ 1, then decreases to γ ≈ 0.8 in 2009; then it increases again in ascending phase of the 24th solar activity cycle. The expression $R^{\gamma(t,R)}$ is larger in maxima epoch than in the minima epoch. On the other hand diffusion coefficient $K_{II}$ should be larger in minimum epoch than in maximum epoch of solar activity. It is adjustable by the changes of a rate of the function $K(t)$ in the manner that the product of $K(t)K(R,\gamma(t,R))$ should diminish from minimum (1996) to maximum (2000) epochs of solar activity and then increase again (Fig. 9) This approximation of diffusion coefficient agrees with the formulation presented e.g. in [Hedgecock, 1975]. The incorporated in the model changes of diffusion coefficient for rigidities R=15GV presents Fig. 9.

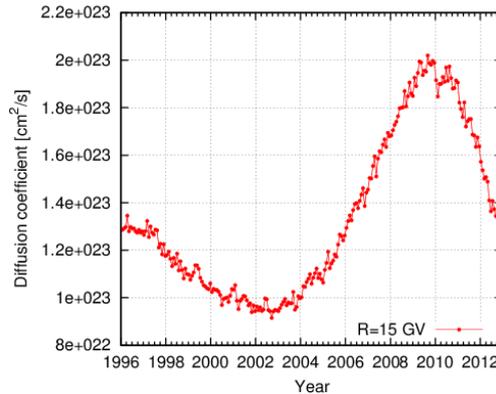

Fig 9. The changes of the diffusion coeficient given by eq. (3) for GCR particles with rigidity 15 GV at the Earth orbit incorporated in the model.

The waviness of the HNS is implemented using the formula of Jokipii and Thomas [1981] $\theta' = \frac{\pi}{2} + TA(t)\sin\left(\frac{r\Omega}{U}\right)$. The temporal changes of tilt angle TA(t) of the HNS incorporated in the modeled period are the observed data from Wilcox Solar Observatory presented in Fig.1.
A drift on the HNS was taken into account according to the boundary condition method [Jokipii and Kopriva, 1979], when the delta function at the HNS is a consequence of the abrupt change in sign of the IMF. Drift effect due to gradient and curvature of the regular IMF is implemented in the model by means of the ratio of the drift $K_d$ diffusion coefficients to the parallel $K_{\parallel}$ diffusion coefficient $\beta_1 = K_d / K_{II}$. In this model we consider that a drift effect is scaled down during the reversal time (2000-2003) of the Sun's global magnetic field considering this period as a complete diffusion dominated period.

Role of delay time between changes of the I and TA is included to transport equation indirectly using temporal changes of diffusion coefficient $K_{II}$. As mentioned above, beside the direct roles of diminishing U and B in the increasing of I, the maximum intensity I in 2009, is established by drift effect with delay time 5 BRs between I and TA in negative polarity (A<0) period. For A<0 a maximum I of the GCR intensity in heliosphere is observed immediate proximity to the HNS. So, higher intensity of GCR at the earth orbit corresponds to the location of HNS in ecliptic region, while lower level of the GCR intensity- to location of HNS out of ecliptic region. Analyzing the dependence between I and TA (Fig. 5), we have found that diffusion coefficient $K_{II}$ approximately linearly increases from 2008 up to middle period of 2009 including role of decreasing U and B. An increase of $K_{II}$ by 30% (Fig. 9) results in a satisfactorily compatibility of solutions of the transport equation with neutron monitors data observations.

The equation (2) was transformed to the algebraic system of equations using the implicit finite difference scheme, and then solved by the Gauss-Seidel iteration method using the boundary conditions $f|_{r=100AU} = 1$, $\frac{\partial f}{\partial r}|_{r=0} = 0$, $\frac{\partial f}{\partial \theta}|_{\theta=0} = \frac{\partial f}{\partial \theta}|_{\theta=\pi} = 0$, and the initial condition with respect rigidity $f|_{R=100GV} = 1$ and with respect the time $f(r,\theta,R_k,t)|_{t=0} = f(r,\theta,R_k)$. The solutions for each layer of rigidity $R$ (R=100, 90, 80,....,10 GV) for the stationary case are considered as an initial conditions for the nonstationary case for the given rigidity $R$ and at time $t$=0. The equation (2) in spherical coordinate system for dimensionless variables is derived in detail in [Siłuszyk et al. 2011], while the details of its numerical solution of the 3D nonstationary equation are given in [Wawrzynczak and Alania, 2010b]. Results of the numerical solution of the equation (2) for rigidity R=15 GV (blue line) in comparison with the changes of the GCR intensity observed by Oulu NM recalculated to the heliosphere presents Fig.10. Fig. 10 demonstrates that there is better consistence between experimental data (Oulu NM) and expected from modeling results for 2004-2012 (A<0) including an increase of the GCR intensity in 2009, while for 1996-2003 the agreement is smaller. Thus, we state that an increase of the GCR intensity in 2009 is caused by diminishing of the solar wind velocity U and the strength B of the IMF, and by drift in A<0 epoch of solar minimum 23/24.

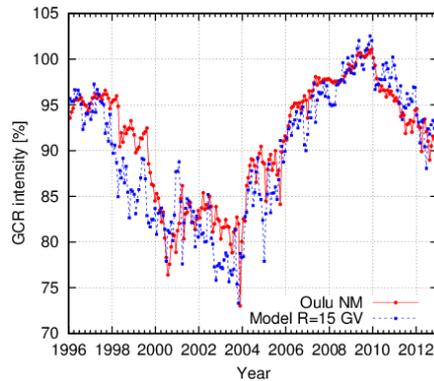

Fig. 10. The GCR intensity variation for the Oulu NM (red solid line) and expected from modeling (blue dashed line) results for the period of January 1996 – December 2012

## 5. Discussion and conclusion

1. In the descending epoch 2004-2010 of solar activity #23 are plainly visible four (11/2003-5/2004, 8/2005-3/2006, 1/2007-5/2007 and 6/2008-5/2009) clearly step-like increases of the

GCR intensity I (according to NMs data) having the plateaus type shape lasting ~6-8 months. Four such events for one half (A<0) of the descending 11-year cycle #23 of solar activity 2004-2010, is the largest number of step-like changes of the GCR intensity observed for NMs era. These features can be associated with the creation of the almost constant homogeneous electro-magnetic conditions in the restricted local vicinity of the heliosphere, where a formation of the impulsive step-like changes of the GCR intensity takes place. Consequently, the heliosphere in the considered period can be regarded as the dynamic structure of concentric sphere-like zones with homogeneous, nearly constant electro-magnetic conditions (intensity of GCR in scope of each sphere-like zone remains almost constant). These unique step-like changes of the GCR intensity in 2003-2009 during descending 11- year solar cycle (A< 0) needs a further study in view of a mechanism of the 11-year variation of GCR [e.g. Lockwood and Webber, 1996]. The step-like changes of the GCR intensity are observed preferentially in the A<0 polarity epochs, when the MIRs and GMIRs type structure are preferentially observed in the heliosphere [e.g., Burlaga and Ness, 2000].

2. Studying relationships between changes I & U, I & B, and I &TA (Fig. 3 - Fig. 5) in the period 2008-2010 we showed the existence of: (1) an anti-correlation between I & U with the maximal magnitude of correlation coefficient (r=-0.68±0.09) for zero delay time (number of BRs is equal to zero); a decrease of U causes a reduction of convection and an increase of I; (2) an anti-correlation between I & B with the maximal magnitude of r (r=-0.73±0.08) for delay time ~ 3 BRs; a dropping of B causes an increase of diffusion coefficient, and consequently, an increase of I, and (3) a strong inverse correlation between I & TA with correlation coefficient r=-0.92±0.05 for the delay time ~5 BRs. Beside the direct roles of diminishing U and B in the increasing of I, the maximum intensity I in 2009 is established by drift effect with delay time 5 BRs between I and TA in negative polarity (A<0) period; for A<0 polarity epoch a maximum I of the GCR intensity in heliosphere is observed immediate proximity to the HNS's with delay time 5 BRs. So, higher intensity of GCR at the earth orbit corresponds to the location of HNS in ecliptic region, while lower level of the GCR intensity- to location of HNS out of ecliptic region, also for delay time ~ 5BRs.

3. For further studding features of solar minimum 23/24 we calculated rigidity dependence of the GCR intensity changes dI before and after 2009. Assuming power law character of the rigidity R dependence of the GCR intensity variations dI (dI~$R^{-\gamma}$) we showed that the exponent γ is gradually diminishing (from γ= 1± 0.07 in 2007, to γ=0.77 ± 0.08 at the end of 2008) before reaching the maximum intensity in 2009. After the 2009 the exponent γ gradually increases (from γ= 0.69 ± 0.09 at the beginning of 2010 up to γ=1.28± 0.12 at the end of 2012). We can conclude that near minimum epoch of solar activity 2009 as for ascending and descending sides of intensity I (A<0) a rigidity spectrum is hard as it is observed generally for few minima epochs of solar activity [Alania et. al. 2008, 2010].

4. Composition of a relatively realistic model of GCR propagation describing long period variations in heliosphere is not easy due to necessity of setting up a few free modulation parameters. To account for an increase I in 2009 we attempted to use parameters which are measurable and estimable proxies. We implemented into transport equation the 27-day averages of the in situ measurements of B, solar wind velocity U and TA. The B influence on propagation of cosmic rays through diffusion coefficient K, thus that is natural to assume a dependence as follows, K∝$B^{\alpha}$, α is unknown. To find α we have built a scattering diagram for dependence of I on B in 1996-2012 (fig .8). We found that α=-0.26±0.03. So, role of B in diffusion coefficient is reduced comparing with case α=1, which is commonly used.
5. We composed relatively realistic 2-D non stationary model to describe variations of the GCR intensity during 1996-2012 including an increase in 2009. We obtained results compatible with NMs data, especially in 2009 (A<0), when diffusion coefficient of cosmic rays for rigidity 10-15 GV was increased by ~ 30% comparing with 1996 (A>0).
6. We conclude that based on investigations presented in this paper there is not possible to find any argument to assume that processes of modulation were unusual during the uncommon prolonged solar minimum 23/24 for GCR to NM are respond.

**Acknowledgements.**

Some parts of this paper were discussed with Prof. H.S. Ahluwalia. We used Neutron Monitors, OMNI and Wilcox Solar Observatory data resources.
A.W. work is supported by The National Center for Science grant awarded by decision number DEC-2012/07/D/ST6/02488. R.M. work is supported by Polish Ministry of Science, Funding for Young Scientist, number 7913/MN.
We would like to thank the reviewers for helpful suggestions.